# Non-trivial band topology in the Superconductor AuSn$_4$: A first principle study

N. K. Karn[1,2], M. M. Sharma[1,2] and V.P.S. Awana[1,2*]

[1]Academy of Scientific & Innovative Research (AcSIR), Ghaziabad, 201002, India

[2]CSIR- National Physical Laboratory, New Delhi, 110012, India

**Abstract:**

Topological semimetals such as Weyl or Dirac semimetal with superconductivity have emerged as a new class of topological materials to realize and study Majorana Fermion. This article reports the Density Functional Theory (DFT) calculated bulk electronic band structure of recently discovered topological superconductor candidate AuSn$_4$. The study has been performed on AuSn$_4$ considering two space groups symmetries viz. Aea2 and Ccce as reported earlier. This study is further extended to the calculation of Z2 invariants. The Fermi surfaces corresponding to the bands, which are responsible for non-trivial band topology along with the surface states are also mapped. The complete study suggests that AuSn$_4$ is a topological semimetal. On AuSn$_4$, it is the first report in the literature showing the non-trivial band topology based on first-principle calculations.

*Corresponding Author

Dr. V. P. S. Awana:  E-mail: awana@nplindia.org

Ph. +91-11-45609357, Fax-+91-11-45609310

Homepage: awanavps.webs.com

**Introduction:**

Topological materials have revolutionized the field of condensed matter and with the discovery of topological materials, some new phases of matter have emerged, and in-depth progress has been made for their classification [1-4]. Generally, materials are put into three categories insulator, semiconductor, and conductor. Since the advent of topology in materials, they have been further classified as topologically trivial and non-trivial [5]. Topological insulators have a bulk bandgap; however, this gap is nonequivalent to traditional insulator vacuum due to surface state bands [6]. The topological semi-metal and metal have no bulk bandgap, and their valance and conduction bands cross within the Brillouin zone (BZ), yet Dirac-like band crossings appear in momentum space, which are protected by discrete symmetries [7,8]. Owing to low energy fermionic excitations and underlying symmetries present in the system, the energy band-crossings can have several-fold degeneracies [9]. The materials with two-fold and four-fold degeneracy points are classified as Weyl and Dirac type, respectively [10]. Beyond these, materials exhibiting three-fold and six-fold degeneracy points are also reported to have non-trivial topology [9, 11-13]. In addition to such point degeneracies, it is possible for a topological material to have band degeneracy along a high symmetry line extending to end of BZ, these types of topological materials are regarded as Nodal Line semimetal. The band degeneracy may also exist along a closed path within the BZ forming degenerate band loops, known as Nodal loop semimetals [14]. Further, topological materials also classified as high symmetry point semimetals (HSPSM) or high symmetry line semimetal (HSLSM), this classification is based on the band degeneracy whether it is lying on high symmetry points of BZ or along high symmetry line.

Numerous methods have been implemented to understand and explain physics associated with the topological materials, including experimental techniques ARPES [15] and transport studies [16]. Such materials show peculiar properties like colossal magnetoresistance [16, 17], high surface conductivity [13, 18]. Among theoretical methods, specific Hamiltonian modeling, the *kp* method [19], DFT-based first principle methods, and others have been developed to study the features of topological materials [2, 20]. In topological materials, the topological surface states are protected by certain symmetries viz., time-reversal symmetry (TRS), reflection symmetry, and accordingly these are characterized by corresponding topological invariants – $Z_2$, and mirror Chern number [21,22].

The superconductors have a bulk superconducting gap, whereas the topological materials have conducting topological surface states. The unique combination of both properties in a single material, makes the same an excellent candidate to host Majorana fermions [23, 24]. Apart from such materials, various interfaces and hybrid devices such as superconductor-semiconductor junction [25], 1D nano-wire in the proximity of s-wave superconductor [26, 27] show the existence of Majorana zero modes. For hybrid devices, one element is superconducting, while the other is chosen to have sufficient spin-orbit coupling (SOC) to give rise to the topological effect [28, 29]. The superconductivity in the bulk of topological material can also be induced either by doping of suitable elements [30-32] or by applying high pressure [33].

In the present article, the spotlight is on the first principle topological study of $AuSn_4$, which is recently found to show some glimpse of superconductivity with topological non-trivial character [34]. Superconductivity in $AuSn_4$ was discovered a long time ago [35], but the system remained unexplored till date. $AuSn_4$ is isostructural to some recently found NLSMs viz. $PtSn_4$ [36] and $PdSn_4$ [37]. Topological semimetals provide greater opportunity to explore superconductivity along with topological character [38,39]. Topological properties of superconducting $AuSn_4$ are not extensively studied till date. There is one theoretical report available in literature that suggests $AuSn_4$ to be a HSPSM, that too in a catalogue of topological materials. Other first principle study of $AuSn_4$, is done for physical properties like mechanical and thermal properties [40]. The system $AuSn_4$, lacks a dedicated theoretical study on its topological properties and the same is warranted in ref. [34].

In this work, we study orthorhombic system $AuSn_4$ using the first principle methods to explore the topology present in the system, following the initial experimental evidence of non-trivial topological surface states [34]. Theoretical calculations of $AuSn_4$ are performed by considering two independent space groups viz. Ccce and Aea2, however the obtained results are similar for both of the space groups. It is found that $AuSn_4$ possesses non-trivial topological character with nodal line like features in the electronic band structure. In-depth analysis of band structure and density of states (DOS) with (w) SOC and without (w/o) SOC in a matrix $6\times6\times5$ has been made to get more insight about the non-trivial topological character of $AuSn_4$. Further, the presence of topology in $AuSn_4$ is also verified through the calculation of Z2 invariants.

**Methodology:**

The first principle-based DFT is implemented in Quantum Espresso [41, 42] to optimize the crystal structure and self-consistent functional calculations (SCF) have been performed to obtain the ground-state electron density and wavefunction. The Perdew-Burke-Ernzerhof (PBE) type generalized gradient approximation (GGA) was used to account for the electronic exchange and correlation [43]. The wave functions were expanded in a plane wave with a charge cut-off of 480 units and wave function cut-off 60eV on a Monkhorst-Pack k-grid of 6×6×5. This computation produces the bulk electronic band structure and Density of states (DOS) of material $AuSn_4$. To find any effect of magnetism on bands, DFT+U (Hubbard correction) calculation are implemented. Further, the Bloch wave-functions generated by DFT are wannierized using WANNIER90 software [44] which produces maximally localized Wannier functions (MLWFs). Wannier functions are the representation of Bloch wavefunction in real space, whereas the MLWFs are Wannier functions with minimized spread with respect to the gauge choice [44]. We reproduced the electronic band structure and matched it with the result obtained by the direct DFT method to check the quality of obtained MLWFs. Out of 92 Bloch wave-functions, 44 are wannierized and disentanglement calculation converges with threshold limit $10^{-10}$. The band structure is reproduced in the energy range ± 2 eV around fermi level ($E_F$), by selecting projectors s,p orbitals of Sn and s, d orbitals of Au atoms. Moreover, an effective tight-binding model is obtained based on these MLWFs for the material under investigation, $AuSn_4$. For topological properties and calculation of Z2 invariant, the tight-binding model is post-processed in software WANNIER TOOLS [45]. The Wannier charge centers (WCC) are obtained of whose evolution in Brillouin zone planes indicated the states of Z2-invariant. The whole Brillouin Zone is sampled on a 6×6×5 K-mesh throughout the computation.

**Results and Discussion**

**Crystal Structure:**

The crystal structure of $AuSn_4$ is orthorhombic, but the space groups are found to be Aea2 and Ccce space group. The most of the old as well as recent experimental results show $AuSn_4$ to have Aea2 space group symmetry [33, 34, 46, 47]. However in online database the space group is found to be Ccce [48,52]. In both cases, the unit cell and the locations of the Au and Sn atoms are same as shown in figure 1(a). Au atoms are at the corners of the unit cell as well as on all face centers while Sn atoms form double layers in between the Au atoms layers. The conventional unit cells for both space groups are same as shown in figure 1(a) but the crystallographic axes are different. All of the literature agrees that it has an orthorhombic structure, but two different space groups are reported as discussed above. Thus, we perform the first principle calculation with both the space groups viz. Aea2 and Ccce. The symmetrized unit cell parameters were taken from the materials project database [48] for Ccce space group are $a = 6.682(3)$Å, $b = 11.871(5)$Å, $c = 6.658(2)$Å and $\alpha = \beta = \gamma = 90^0$. For Aea2 type space group, experimentally obtained unit cell parameters were obtained from a very recent experimental report on the studied system [47]. The cell parameters and atomic positions for both cases are shown in table 1 and table 2 respectively. In order to visualize any difference between the two structure, theoretical powder XRD intensity patterns are obtained and compared. The calculated powder XRD intensity plots are shown in figure 1(b). All peaks match with each other except the relative intensity heights. Though both of the space group seems to be similar, yet they are different from the point of symmetry, as Ccce is centrosymmetric whereas Aea2 is non-centrosymmetric.

Figure 2(a) and 2(b) represents the first Brillouin zone for both the space groups viz. Ccce and Aea2 respectively, which are drawn using visualization software Xcrysden [51]. Interestingly, the first Brillouin zone of both of the space group is hexagonal, but the reciprocal axis differs due to different crystallographic axis. The crystallographic axes difference is also evident from the single crystal XRD peaks of isostructural materials with different space groups. The single crystal X-Ray Diffraction (XRD) pattern contains the peaks for $(002n)$ planes for space group Aea2 [34], while for Ccce space group the observed peaks are from $(02n0)$ [37].

**Band Structure:**

The unit cell parameters for AuSn$_4$ are taken from the online database for Ccce space group [48] and for Aea2 space group the same is taken from ref. [47]. Structure optimization calculations are performed within the etiquette of Density functional theory (DFT) for both the phases. The bulk electronic band structure and density of states (DOS) are computed based on optimized structure. Electronic band structure and DOS are calculated by considering with (w) SOC and without (w/o) SOC effects. The electron density change is approximated by gradient functional – the generalized gradient approximation (GGA type functional). The high symmetric point path in the first BZ is obtained as Γ → S→ R → Z →T. The first BZ and the calculated path for electronic band calculation for both the space groups viz. Ccce and Aea2 are shown in figure 2(a) and 2(b) respectively. In the Figure 3(a), the upper image shows the calculated bulk electronic band structure along with the DOS just below it for the Ccce space group structure, while figure 3(b) shows the same for Aea2 space group. The whole band is shifted so that the fermi–energy can be set at zero as shown in the plots. The band structure calculation reveals that the effect of SOC is not very large. However, the zoomed images shown in figure 4, tells that SOC is not completely redundant. The presence of line degeneracy is evident in w/o SOC plot along S → R path in figure 4(a) and 4(b), which signifies the presence of nodal line structure. The line degeneracy along the path S → R is lifted, while the degeneracy at the high symmetric points S and R remains intact for both space groups, as shown in figure 4(a) and (b). The double Dirac cone type structure appears for both space groups when SOC parameters are not included as shown in figures 4(c) and 4(d). When SOC is included, the gap opens up in both cases and creates a possibility of band inversion. Interestingly, the splitting of bands around double Dirac cone type structure for both the space groups are not identical as shown in figure 4(c) and 4(d). The double degeneracy appeared in w/o SOC band structure has been lifted with inclusion of SOC parameters for both of the space groups. In case of Ccce space group, the degeneracy at Dirac point is lifted in the form of two double degenerate bands as shown in figure 4(c). In case of Aea2 space group, at the same place, wSOC band split into 4 non degenerate bands in contrast to Ccce space group where two gapped double degenerate bands appeared. The difference in band splitting for both of the space can be attributed to inversion symmetry present in the Ccce space group, which is absent in Aea2 space group. Another notable difference between the two space group bands is that a single Dirac cone is present in band structure for Aea2 space group along path R-Z near Fermi-level at -0.5eV. Such Dirac cone is missing in the band structure plot calculated for the Ccce space group.

The DFT many times underestimates the band gaps. To avoid such possibility, we performed the DFT+U calculation which uses Hubbard corrections. The SCF calculations was done with taking non-zero initial magnetization. It converged with zero magnetization for both atoms. Next, Hubbard parameters (U) for Au is 4eV and the same for Sn is 3.7eV is taken from [52] and including those corrections DOS and Band structures were calculated. No significant change is observed in the electronic band structure calculated using DFT+U formalism, also DOS is non-zero at the fermi level. The same results are also reported in AFLOW database [52] for AuSn$_4$.

The fat-band analysis for AuSn$_4$ bands in WANNIER90, shows that the bands near the Fermi level have significant contributions from s, d-orbitals of Au and s, p-orbitals of Sn atoms. Non-zero DOS at the Fermi level confirms the metallic/semi-metallic behavior of AuSn$_4$, and there are five bands that cross the fermi-energy. Figure 5(a) and (b) show the calculated Fermi surface of AuSn$_4$ corresponding to the bands crossing the crossing fermi level. For Ccce space group there are three fermi surfaces, which cross the first BZ, whereas for Aea2 space group only two fermi surfaces cross the first BZ.

**Surface States**:

The overall results for both of the space groups viz. Ccce and Aea2 are found to be identical, while Ccce space group preserves more symmetry as compared to Aea2 space group. Taking into account the similarity between the two space groups and the symmetry present in Ccce space group, the Fermi energy surface is calculated by projecting it into the 2D plane (001) for the Ccce space group. Figure 6(a) and (b) show the calculated surface states in $k_x k_y$-plane w/o SOC and w SOC, respectively, here the Γ point is taken as the center of the plane. The bands for which the energy dispersion is calculated are marked as 1 and 2 in figure 3(a). The band marked 1 corresponds to the upper energy dispersion and the band marked 2 corresponds to the lower energy dispersion. There is a degeneracy at the center in energy dispersion calculated without considering SOC parameters which is further surrounded by several line degeneracies. The degeneracy at the center is shown in the figure 6(c) by taking a vertical cut of figure 6(a) in yz-plane. The line degeneracy can be easily seen in the gap plane calculation shown in Figure 7(a). The difference between the energy dispersion of the two energy surfaces is plotted and the same is shown in figure 7(a). These degeneracies are lifted when the SOC is included in the calculation. The splitting of the degeneracy at the center is encircled in Figure 6(d). The vanishing line degeneracies can be easily seen as vanishing dark color maps, in the gap plane calculation as shown in Figure 7(b). From the same plot we find that, at central point degeneracy, a gap of 0.15eV is induced. This analysis gives clear indication that the system under investigation has topological character and with non-vanishing DOS at the Fermi level it can be classified as topological Dirac type-I semimetal. Further, the energy dispersion gap is also calculated for Aea2 space group in with SOC case as shown in figure 7(c). The whole spectrum is found to be gapped out which is similar to Ccce space group.

**Z2 – invariant**:

Recently a series of theoretical developments has allowed us to classify and quantify the topology present in the system by evaluating Z2-invariants. Different Z2-states are separated by a topological phase transition which involves an adiabatic transformation of the bands leading to the gap closing of the two bands. There have been several methods developed to calculate Z2 invariant - discrete method involve finding Pfaffians/parity over the fixed points of the time reversal symmetry [53], Fukui- Hatsugai method [54] and by tracking the windings of Wannier Charge Center (WCC) evolution around the BZ [55,56]. Here we adopt WCC method to determine Z2 invariant which has been adjoined with WANNIER TOOLS for numerical implementation. The WCCs are the pseudo charge points. Their locations are the extrema points of probability distribution of MLWFs, locations of maximum probability of MLWFs regarded as negative charge and with least (zero) probability regions are regarded as positive charge. Exchange of these charge centers in different k-planes define trivial and non-trivial topology. So from the tight binding model obtained for Wannier functions, the WCC are computed numerically and evolved in the 6-planes – $k_x$, $k_y$, $k_z = 0$ and $k_x$, $k_y$, $k_z = 0.5$ in the Brillouin zone. An odd number of the crossing of WCC implies a topologically non-trivial state (Z2=1), whereas an even number of crossings indicate the presence of a topologically trivial state (Z2=0). The w/oSOC bands have degeneracies for both type of space group but the wSOC bands are gapped out which enable us to calculate Z2 invariant for wSOC bands. Table 3 shows the results for Z2 invariant of numerical implementation for both type of space groups for wSOC bands.

For the 3D systems the Z2 topological invariant is represented by four index ($v_0; v_1v_2v_3$). The first index designate strong topology calculated as the $sum\ of\ Z2\ in\ k_i\ plane\ mod\ 2$. The rest three index given by $v_i = Z2\ value\ in\ (k_i = 0)\ plane$, indicate a weak topology present in the system. The strong index has some redundancy as one can see from the Table 3, for Aea2 space group, in $k_yk_z$ plane the strong index is one i.e. non-trivial. But for the remaining plane $k_xk_y$ and $k_yk_z$ the strong index is zero – trivial for both cases. For Ccce also, the strong index has redundancy as in $k_xk_z$ and $k_yk_z$ plane it is one but in $k_xk_y$ it is zero. Here the weak index is (010) for Aea2 and (111) for Ccce space group. Clearly, with SOC calculations dictate the presence of weak topology in both type of the crystal system. From Table-3, it is found that the two phases of AuSn$_4$ with different space group are topologically distinct phase as the Z2 value completely changes from 0 to 1 or vice versa in $k_xk_y$ and $k_yk_z$ plane for both the space groups. However, for $k_y$=0.5 plane the Z2 value does not change. Thus, Z2 invariant analysis confirms that topological character present in the system but topologically the two phases are distinct.

**Conclusion:**

To summarize, the optimized structure of AuSn$_4$ in two phases (Aea2 and Ccce) is computed and corresponding to that the electronic band structure and DOS were calculated with DFT for both case w/o SOC and with SOC. The analysis of the same confirms both phases to be semimetallic with Dirac type-I nodes. The SOC appears to be dormant but energy dispersion analysis confirms the same to be effective, which further removes the line degeneracy. The Z2 invariant calculation confirms the weak non-trivial band topology present in both phases of studied system. However, the two phases are topologically distinct as confirmed by the Z2 invariant evaluation of the two phases. Theoretically studied topological properties along with the observed superconductivity in AuSn$_4$, establishes the same as a good candidate of topological superconductor. This is the first ever detailed theoretical report on topological properties of AuSn$_4$, which will certainly open door for further advancement in the field.

**Acknowledgement:**

The authors would like to thank Director NPL for his keen interest and encouragement. N.K. Karn and M.M. Sharma would like to thank CSIR for the research fellowship. Both the authors are also thankful to AcSIR for Ph.D. registration.

**Conflict of Interest statement:**

Authors have no conflict of interest.

**Figure Captions:**

**Figure 1(a):** Orthorhombic crystal structure of AuSn$_4$ as generated by VESTA. Left – conventional uni cell for Ccce space group, Right – conventional unit cell for space group Aba2.

**Figure 1(b):** XRD pattern generated by VESTA for both of the optimized crystal structures.

**Figure 2:** Shows the first Brillouin zone with the coordinates of high symmetric points. Green arrow shows the path chosen for the Band structure calculation. (a) For Ccce space group (b) For Aea2 space group

**Figure 3:** The calculated bulk electronic band structures shown for both space group along with the Density of States(DOS) w/o and with SoC within the protocols of Density Functional Theory(DFT). (a) For Ccce space group (b) for Aba2 space group

**Figure 4:** Zoomed view of bands near the Fermi level. (a)For Ccce space group upper image show the nodal line structure vanishing due to SOC. Lower image show the Dirac cone type features present in the band structure. (b) Very similar features present for the Aba2 space group is shown.

**Figure 5:** Fermi surface is shown corresponding to the five bands crossing the Fermi level. (a) For Ccce space group (b) For Aea2 space group

**Figure 6:** Surface energy dispersion in plane (001) for the Ccce phase of $AuSn_4$. (a) and (b) show energy dispersion w/o and with SOC. (c) The degeneracy at the center for w/o SOC is shown in the red encircle. (d) Shows central degeneracy vanishes by including SOC

**Figure 7:** Gap in energy dispersion planes shown with color gradient. For Ccce phase (a) w/o SOC, shows central degeneracy is surrounded by line degeneracy (b) With SOC shows how line degeneracy vanishes and central point degeneracy is also lifted. For Aba2 phase (c) shows the gapped energy dispersion plane for wSOC bands.

**Figures:**

**Figure 1(a)**

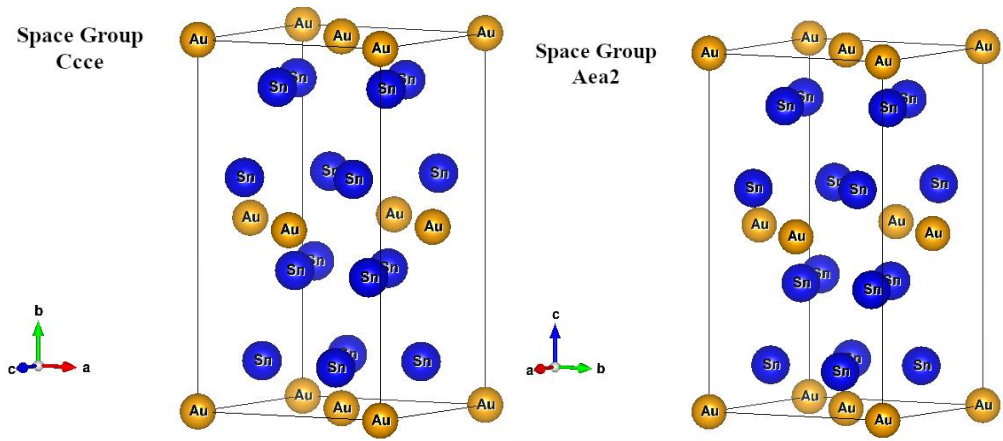

**Figure 1(b)**

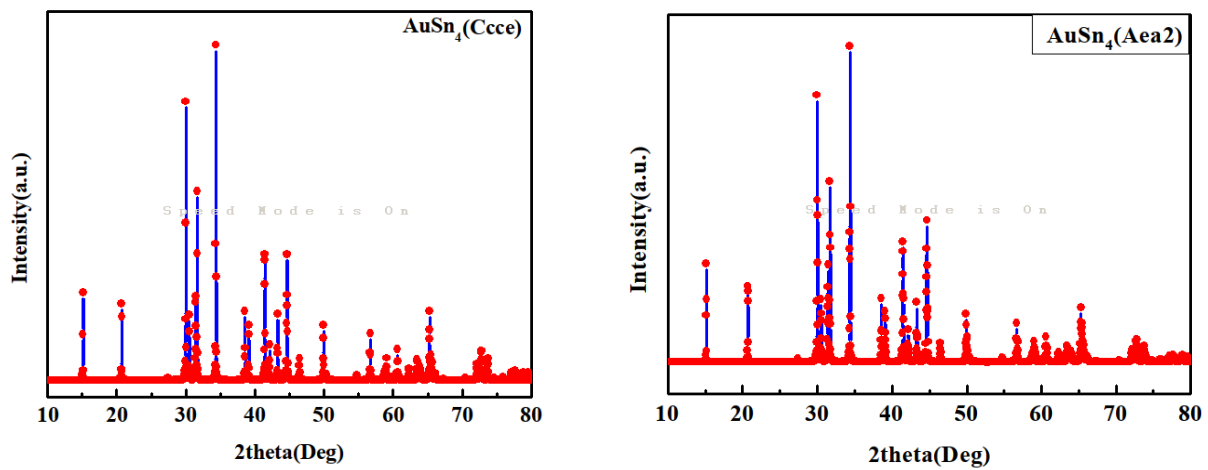

**Figure 2(a)**

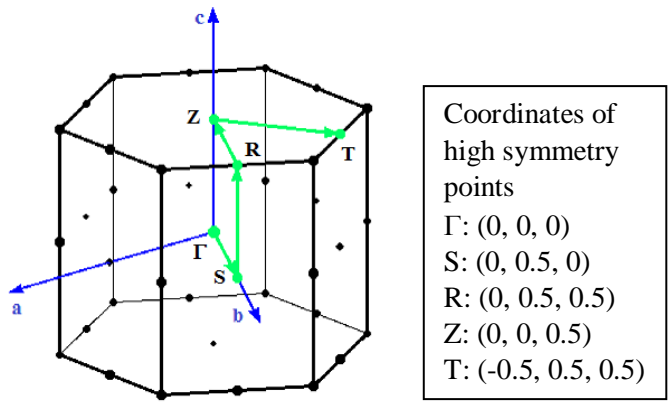

Coordinates of high symmetry points
Γ: (0, 0, 0)
S: (0, 0.5, 0)
R: (0, 0.5, 0.5)
Z: (0, 0, 0.5)
T: (-0.5, 0.5, 0.5)

**Figure 2(b)**

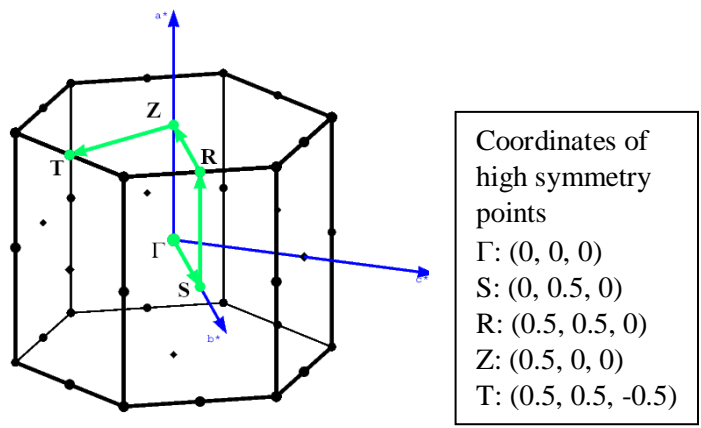

Coordinates of high symmetry points
Γ: (0, 0, 0)
S: (0, 0.5, 0)
R: (0.5, 0.5, 0)
Z: (0.5, 0, 0)
T: (0.5, 0.5, -0.5)

**Figure 3(a)** **(b)**

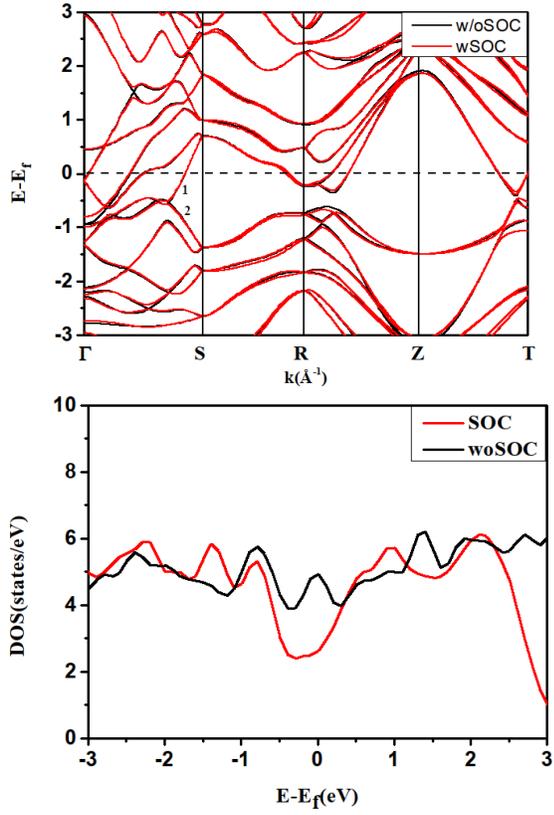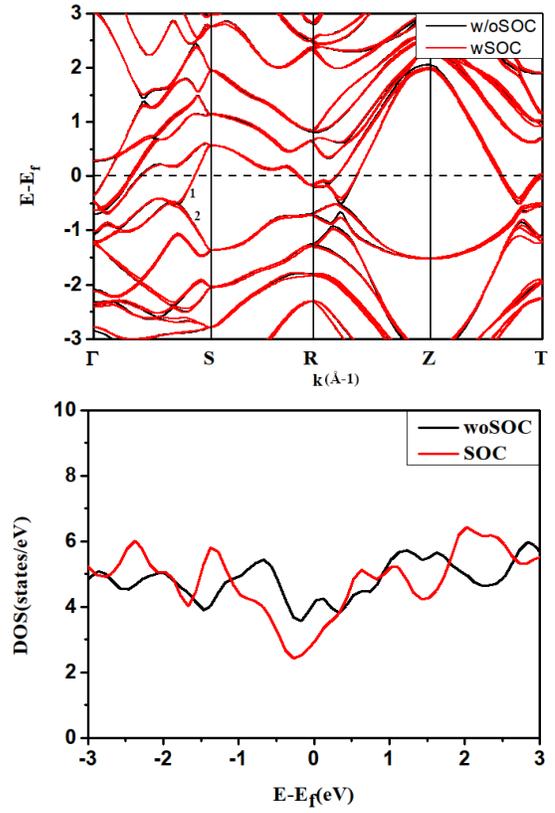

**Figure 4(a)** **(b)**

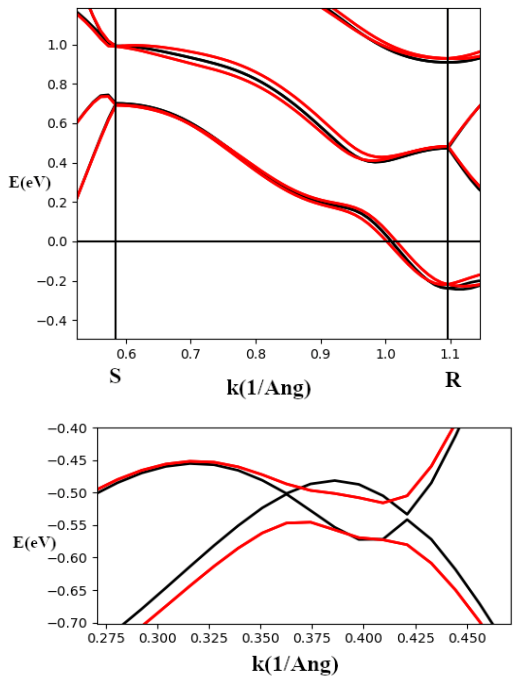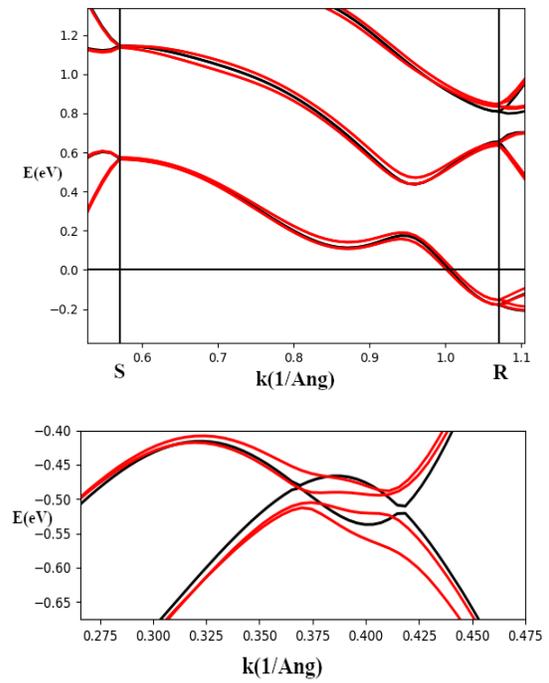

**Figure 5(a)**

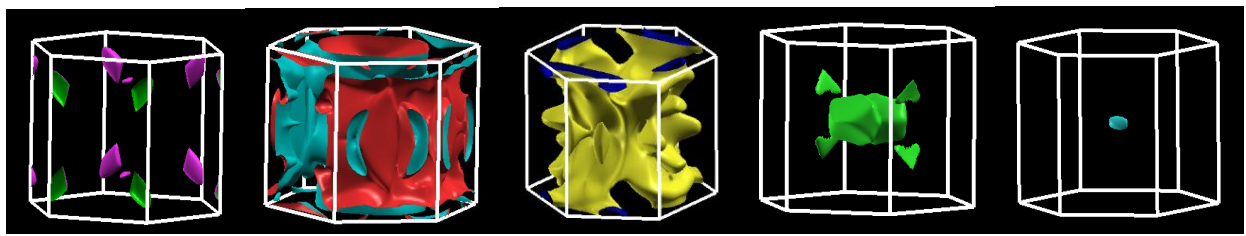

**Figure 5(b)**

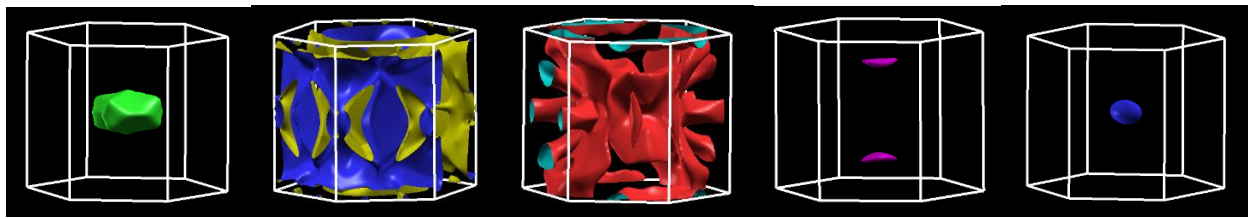

**Figure 6(a)**

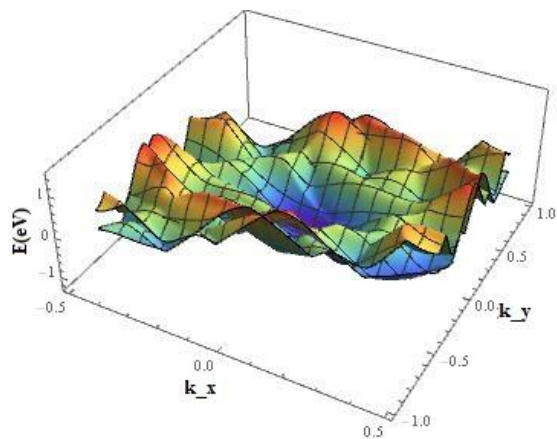

**Figure 6(b)**

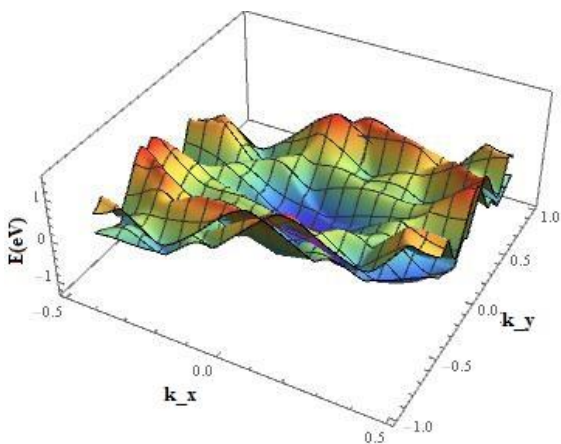

**Figure 6(c)**

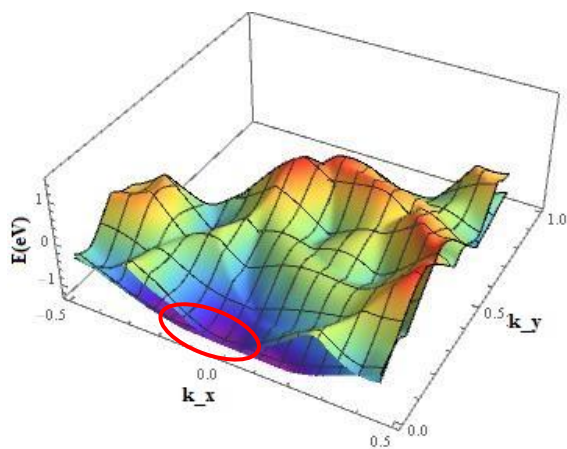

**Figure 6(d)**

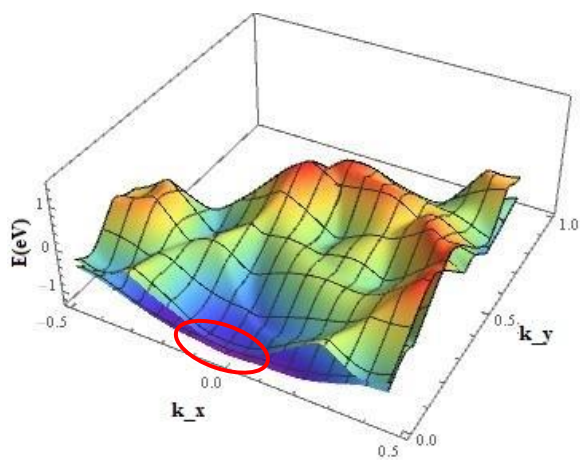

**Figure 7(a)**

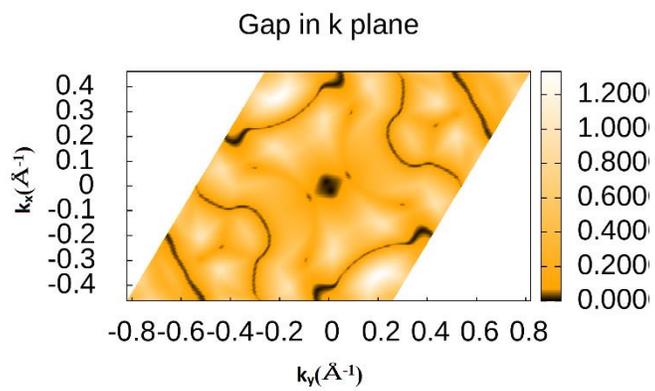

**Figure 7(b)**

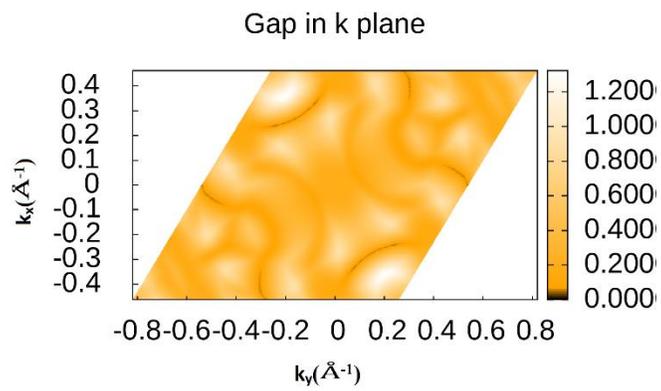

**Figure 7(c)**

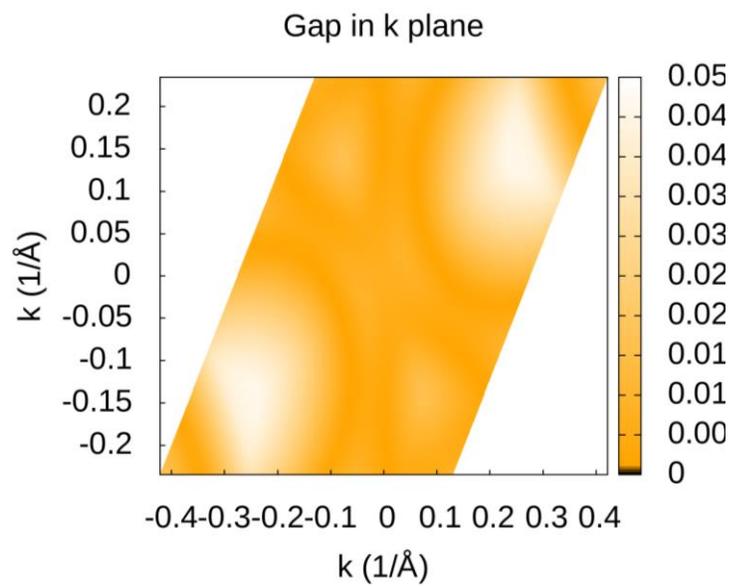

**Table 1 : Structural Parameters for AuSn$_4$**

| Space Group<br>Lattice Parameters | Aea2 | Ccca |
|---|---|---|
| a | 6.503(6)Å | 6.682(4)Å |
| b | 6.518(6)Å | 11.871(6)Å |
| c | 11.7173(12)Å | 6.658(2)Å |
| $\alpha$ | $90^0$ | $90^0$ |
| $\beta$ | $90^0$ | $90^0$ |
| $\gamma$ | $90^0$ | $90^0$ |

**Table 2 Atomic co-ordinates of AuSn$_4$ for both of the space groups viz. Ccce and Aea2 :**

| Ccce | | Aea2 | |
|---|---|---|---|
| Au | (0,0,0) | Au | (0,0,0) |
| Sn(1) | (0.16390,0.33950,0.12060) | Sn | (0.16007,0.62650,0.83332) |
| Sn(2) | (0.33120,0.16240,0.85310) | | |

**Table 3 Calculated Z2 invariants for considered space groups viz. Ccce and Aea2.**

| Plane | With SOC Z2 (Aea2) | With SOC Z2 (Ccce) |
|---|---|---|
| $k_x = 0.0$, $k_y$–$k_z$ plane | 1 | 0 |
| $k_x = 0.5$, $k_y$–$k_z$ plane | 0 | 1 |
| $k_y = 0.0$, $k_x$–$k_z$ plane | 1 | 0 |
| $k_y = 0.5$, $k_x$–$k_z$ plane | 1 | 1 |
| $k_z = 0.0$, $k_x$–$k_y$ plane | 0 | 1 |
| $k_z = 0.5$, $k_x$–$k_y$ plane | 0 | 1 |